\documentclass[epj]{webofc}
\usepackage{color}
\usepackage[justification=justified,font=small,labelfont=bf]{caption}
\usepackage[justification=justified]{subfig}
\usepackage[varg]{txfonts}   
\wocname{EPJ Web of Conferences}
\woctitle{INPC 2013}
\begin{document}
\title{The european FAZIA  initiative: a high-performance digital telescope array for heavy-ion studies}
%
%

\author{G. Casini\inst{1}\fnsep\thanks{\email{casini@fi.infn.it}} 
\and S.~Barlini\inst{1}
\and G.~Pasquali\inst{1}
\and G.~Pastore\inst{1}
\and M.~Bini\inst{1}
\and S.~Carboni\inst{1}
\and A.~Olmi\inst{1}
\and S.~Piantelli\inst{1}
\and G.~Poggi\inst{1}
\and A.~Stefanini\inst{1}
\and S.~Valdr\'{e}\inst{1}
\and E.~Bonnet\inst{3}
\and B.~Borderie\inst{4}
\and R.~Bougault\inst{5}
\and M.~Bruno\inst{2}
\and A.~Chbihi\inst{3}
\and M.~Cinausero\inst{6}
\and M.~Degerlier\inst{8}
\and P.~Edelbruck\inst{4}
\and J.D.~Frankland\inst{3}
\and F.~Gramegna\inst{6}
\and D.~Gruyer\inst{3}
\and M.~Guerzoni\inst{2}
\and A.~Kordjasz\inst{9}
\and T.~Kozik\inst{10}
\and N.~Le Neindre\inst{5}
\and O.~Lopez\inst{5}
\and T.~Marchi\inst{6,7}
\and P.~Marini\inst{3}
\and L.~Morelli\inst{2}
\and A.~Ordine\inst{11}
\and M.~P\^{a}rlog\inst{5}
\and M.F.~Rivet\inst{4}
\and E.~Rosato\inst{11}
\and F.~Salomon\inst{4}
\and G.~Spadaccini\inst{11}
\and T.Twar\'{o}g~\inst{9}
\and E.~Vient\inst{5}
\and M.~Vigilante\inst{11}
}

\institute{ INFN, Sezione di Firenze and Dipartimento di Fisica,
  Universit\`{a} di Firenze, Italy 
\and  
Dipartimento di Fisica, Universit\`{a} di Bologna and INFN, Sezione di
Bologna, Italy  
\and
GANIL, CEA/DSM-CNRS/IN2P3, B.P.~5027, F-14076 Caen cedex, France
\and 
Institut de Physique Nucl\'eaire, CNRS/IN2P3, Universit\'e
Paris-Sud 11, F-91406 Orsay cedex, France 
\and
LPC, IN2P3-CNRS, ENSICAEN et Universit\'e de Caen, F-14050 Caen-Cedex,
France
\and
 INFN, Laboratori Nazionali di Legnaro, Padova, Italy
\and  
Dipartimento di Fisica, Universit\`{a}  di Padova, Italy  
\and 
Nevsehir University, Science and Art Faculty, Physics Department, Nevsheir,
Turkey
\and
Jagiellonian University, Institute of Nuclear Physics IFJ-PAN,
PL-31342 Krak{\'o}w, Poland. 
\and 
Warsaw University, Poland
\and
Dipartimento di Fisica, Universit\`{a} Federico II di Napoli and INFN,
Sezione di Napoli, Italy 
         }
\abstract{%
The european Fazia collaboration aims
at building a new 
modular array for charged product identification to be employed for
heavy-ion studies. The elementary module of the
array is a  Silicon-Silicon-CsI telescope,
optimized for
ion identification also via pulse shape analysis.  The achievement
of top performances imposes  specific electronics which has been
developed by FAZIA and features high 
quality  charge and current preamplifiers,  coupled to  fully digital
front-end.
During the initial R$\&$D phase, original and novel  solutions have  been 
tested in prototypes, obtaining unprecedented ion identification
capabilities. FAZIA is now constructing  
a demonstrator array
consisting of about two hundreds telescopes arranged in a
compact and transportable configuration. 
In this contribution, we mainly summarize some aspects studied by FAZIA
to improve  the ion identification. Then we will briefly 
discuss the FAZIA program centered on experiments to be done with the
demonstrator.  First
results on the isospin dynamics obtained with a reduced set-up
demonstrate well the performance of the telescope and
represent a good starting point towards future investigations with
both stable and exotic beams.

}
\maketitle
\section{Introduction}
\label{intro}
The studies on the reaction mechanisms occurring in heavy ion
collisions have been recently focused 
on observables related to the isospin\footnote{we adopt  the widespread
usage of isospin for the N/Z ratio} content of the system. In fact,
such observables allow to access the isovector term of the nuclear
potential, whose behaviour is scarcely known for nuclei far from ground-state
conditions. In particular, the density and temperature  dependence of
the nuclear symmetry  energy ($E_{sym}$) are the subject of many
recent theoretical~\cite{baran05,baran12,kumar12} and
experimental~\cite{macin,cardella12,defi12,brown13} studies, which investigated
heavy-ion reactions at different 
regimes of bombarding energy, impact parameter and system size. Since
a common feature of most collisions is the production of multiparticle
final states, a
crucial experimental challenge 
is to detect and recognize
at the best
the many reaction products, as a preliminary necessary
step to trace back to the interaction phases~\cite{dago11}.
Recently, the detector frontier
has become the isotopic identification of fragments in a wide range of
kinetic energies and over large solid angles~\cite{padus05,amorini,henzlo08}. 
Keeping this in mind, some
years ago the european FAZIA collaboration was born from a previous
cooperation among groups involved in this research field at several
laboratories. FAZIA started a research program to optimize the
detectors in terms of isotopic identification and, after
that, to construct a modular array with state-of-the-art performances
that will be used for heavy-ion experiments with both stable and 
radioactive ion  beams, such as those to be delivered by  SPES (LNL),
SPIRAL2 (GANIL) and FRIBS (LNS) facilities.

\section{The FAZIA module}
In order to build a versatile and transportable array for charged
fragments  the FAZIA 
project is based on a conventional solid-state 
telescope as done by
other groups~\cite{padus05,wue09}. The telescope (active area
20x20mm$^2$) is made 
of two Silicon layers (300 and 500~$\mathrm{\mu m}$
thick, respectively) followed by a slightly tapered CsI(Tl)
scintillator 100~mm long.  As usual, this device allows one to
separate particles which punch through a first active layer by means
of the well known $\Delta E-E$ technique, where $E$ is the energy
released in the second 
layer. To increase the dynamical range of the telescope (FAZIA
is supposed to work from 10 to 50~MeV/u), three detectors are used so that
the  $\Delta E-E$ method can be applied using more signal
combinations. In the  $\Delta E-E$ method the energy threshold for ion
identification (Z and/or A) is simply the minimum energy required to punch
through the $\Delta E$ layer. 
A strong effort was done by FAZIA to improve 
the performance 
(separation capability and threshold) while 
maintaining good energy information. Therefore, particular care was
devoted to verifying the limits of identification for particles stopped
in the first silicon, basing on the Pulse Shape Analysis
(PSA). Accurate PSA strongly benefits from digital fast sampling electronics
coupled to high quality analog stages, which preserve the original waveforms.
Indeed, FAZIA complemented the R$\&$D on the detectors with the construction of
high-performing electronic channels, each made of a low-noise
preamplifier with current and charge outputs, followed by complex
stages where the signals are sampled and processed by FPGA to extract
the needed information. Such a concept has been implemented in
the test phases on single channel electronics, while for the
demonstrator array an integrated multifunction FEE  equipment has been
developed which serves many channels and operates under vacuum. \\ 
The results obtained by FAZIA are described in many papers (see
e.g.~\cite{luigi09,luigi09a,luigi11,stefano,nicolas13,barlird}) which
we refer to. In this Section we 
collect an anthology of these results to enlighten some
critical aspects, the proposed solutions and the
obtained performance. Moreover, we introduce some preliminary
results  on the PSA in underdepleted silicons emerging
from the most recent tests.
\label{sec-1}


 \subsection{Channeling effects}
Ion identification via PSA in silicon is critically sensitive to
spurious sources  which modify the native (charge or current)
waveforms. Channeling effects are one of these sources as clearly
demostrated by FAZIA with specific experiments, where a telescope
was remotely and finely rotated under irradiation to select or avoid specific
crystallographic directions~\cite{luigi09,luigi11}. The performance of the
detector strongly degrades for orientations close to
the crystal principal planes or axes: it was
found that energy and charge risetime resolutions worsen by a factor of three
with respect to random orientations. The effect is large not only for PSA
but also for the conventional $\Delta E-E$ method. Accordingly,
FAZIA silicons are made from wafers cut in such a way that the
ions hit them along  'random' paths (about 7$^\circ$ off the <111>
direction) all over the active area. 
\label{sec-2}

 \subsection{Preserving original signal waveforms for PSA}
The time development of signals  in particle detectors depends on the
drift electric field. Hence, for PSA applications, it is important to
avoid any variation of the applied electric field with time and impact
position on the detector, as
pointed out years ago~\cite{pausch97}. FAZIA 
examined this subject using the powerful tools
offered by  digital 
electronics and by  versatile supply commercial equipments.
To reduce variations with time, FAZIA developed
a specific bias supply system  
with a feedback loop which 
measures the current during the experiment and adjusts the 
applied voltage to compensate the 
drop across the bias resistor.\\ 
Concerning the effects due to silicon doping,  
FAZIA studied the  quality of ion   
identification in silicons with different inhomogeneities of dopant concentration.
Indeed, for a given applied junction voltage, the effective
electric field is a 
function of the bulk resistivity (i.e. of the doping) which is therefore
important to keep as constant as possible. With specific
tests~\cite{luigi11},  we
demonstrated the need to keep the inhomogeneities below a few percent
for good PSA applications. No effect, as expected, has been found
on the $\Delta E -E$ method.  
Hence, the present FAZIA recipe requests the use of silicons with high doping
homogeneities, such as those attained nowadays through neutron
transmutation techniques. Our typical  silicons have $\rho =3000-4000~\mathrm{
\Omega \cdot cm}$. 
To verify in a rather easy way, without beam, the
doping homogeneity of silicon crystals, a procedure based on a
UV-laser  has been implemented~\cite{luigi09a},  which allows us to produce a
resistivity map for each detector before its final assembling.\\
We finally mention that PSA is also spoiled by fluctuations induced by
radiation damage effects. The final effect of several types of damage
produced in  
silicon by impinging ions is to cause microscopic variations of the
electric field and/or favour electron-hole recombination or
trapping. As a result, fluctuations of signal shapes increase
and PSA is degraded. Such effects, largely dependent on the ion charge
and energy, are difficult to summarize and to schematize; a
specific experiment to study how PSA degrades with ion fluence has
been done using Xe ions at 35~MeV/u~\cite{barlird}.\\ 
All the above reported effects causing signal shape variations are critical
for PSA, especially when one aims at some isotopic separation which is 
associated to weak variations of the signal shape. In this respect,
once all these spoiling 
contributions have been kept under control (from silicon production and wafer
cut to electronics performances), one can reach a lower threshold for
good~\cite{stefano} mass separation 
$\sim$5~MeV/u~\cite{luigi09} in the region of Carbon ions. 
\label{sec-3}

 \subsection{Front and rear irradiation}
Since the pioneer works on PSA with silicon detectors~\cite{pausch96},
it was argued that reverse mounting operation would improve ion
identification thanks to the overall lengthening of signal risetimes when
ions enter the ohmic side of the device. The choice of FAZIA, from the
very beginning, was indeed to adopt the reverse configuration. 
Possible drawbacks of 
this choice are worse timing and energy resolutions, especially for
slow particles 
stopped in the very low electric field region. As a matter of fact,
in most silicon arrays people adopted the popular ``front''
configuration with the junction (high field) side facing incoming
ions. Recently, thanks to digital electronics and with the goal
to lower ion identification thesholds, the mounting geometry of silicon
detectors has been reconsidered~\cite{alder06,amorini} in order to compare
the resolutions attainable in the two
configurations, but  conclusive tests were
missing. FAZIA studied this aspect with a specific experiment where
a given three-layer telescope featuring high quality silicon stages
(as  described in the previous sections) was used.
The telescope was irradiated in the same controlled
conditions (beam, target, distances and angles, bias voltage and
electronic chain) but in two 
phases: first the silicon chips  have been irradiated in ``front'' mounting,
then they have been flipped and  used for the rest of the
measurement. A comparison of the two obtained data
sets~\cite{nicolas13} 
showed that the $\Delta E-E$ technique gives the very same good
resolution provided that sufficiently long shaping times are used for
signal filtering (to avoid possible ballistic deficit). Sizeable
differences are found, instead, for PSA: front irradiation is
definitely worse. Indeed, of the two employed shape related variables
(plotted in `bidimensional
correlation with the deposited energy E), i.e. the
charge signal risetime (QRT) and the maximum of the current signal
(IMAX), only the first one gives  Z identification in front mounting,
while IMAX fails. In rear mounting, instead, particle are
identified using  both shape variables and, more important, the thresholds
are also improved. By using the E-QRT plot, 
in  front mounting geometry the Z
separation is obtained for ions with range in silicon greater than
170~$\mathrm{\mu m}$ (Z$\le 10$) to 250~$\mathrm{\mu m}$ (Z$\sim 30$, Zinc); in reverse
mounting the thresholds for charge separation drop as low as 30-40~$\mathrm{\mu
m}$ for Z$\le 8$, steadly increasing up to  150~$\mathrm{\mu m}$ for Zinc ions.
\label{sec-4}

 \subsection{PSA in underdepleted silicons: preliminary results}
Preliminary results are reported here on a recent test to investigate
the PSA in underdepleted silicon detectors. So far, the  usual 
operating condition for the FAZIA silicons (as for other
collaborations) corresponded to a moderate (10-20$\%$) 
overbias, in order to  avoid a nominally zero electric field
region and to reduce dead-layers. It was implicitely supposed that
it is a bonus for typical 
\begin{figure}[htb]

\centering
\subfloat[]{\label{fig-1:a} \includegraphics[width=0.5\textwidth,clip]{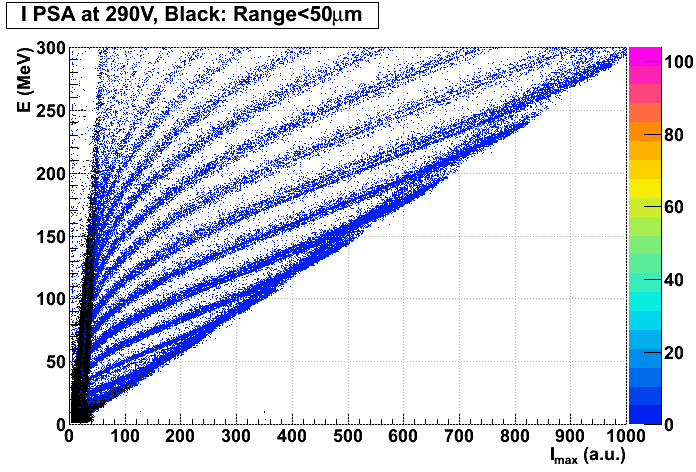}}   
\subfloat[]{\label{fig-1:b} \includegraphics[width=0.5\textwidth,clip]{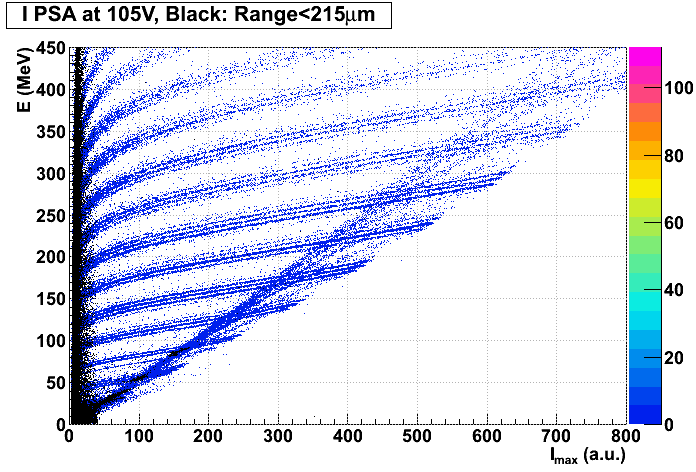}}   
\caption{PSA identification plot of E vs. IMAX for a 500~$\mathrm{\mu m}$ FAZIA
  silicon operated at different voltages. Letf: full depletion
  voltage, 290~V. Ions are separated  in charge but not in mass. Black
  points are events of ions with a range less than 50~$\mathrm{\mu m}$. Right:
strong underdepletion voltage, 105~V. Ions  are separated in charge
and in mass up to Z$\sim$10. Black dots correspond to ions with range
less than an arbitrary value of 215~$\mathrm{\mu m}$. Some punching through
events   appear here, more clearly than at 290~V. They
survive 
the CsI(Tl) veto due to a imperfect geometrical 
match with  the silicon pair.}
\label{fig-1}       
\end{figure}
performance to avoid phenomena
due to an almost vanishing electric field, such as an extremely slow charge
collection or even partial charge carrier deficit (in case of e-h
recombination). However, at least for PSA, the longer
time scales associated to signal formation could in
principle increase the sensitivity and improve the
ion separation. To study this aspect, FAZIA 
recently performed an experiment using a standard silicon telescope
where the first 300~$\mathrm{\mu m}$ silicon was kept at the usual voltage (slightly
overbiased) while the following 500 $\mathrm{\mu m}$ detector was operated at
several voltages, from the nominal depletion (290 V) to a very low
value (105 V), the latter corresponding to an 
undepleted layer (nominal zero field) of 200 $\mathrm{\mu m}$; in this
condition, this second detector could behave as a 300 $\mathrm{\mu m}$ silicon
with an initial undepleted (and possible inactive) layer. 
The first $\Delta E$ silicon is important as a reference to identify
ions via the $\Delta E-E$ method  and to calibrate the raw $E$
axis. Indeed, at underdepletion voltages one expects that the energy
scale can vary with respect to the standard calibration, due to the
presence of a variable dead-layer (we adopt  rear injection
mounting) and, as said, of recombination effects. So, some non-linarity is
expected in the energy scale, also depending on the ion type and
energy. The other expected effect is the increase of the
signal risetimes; indeed, QRT values up to 10~$\mathrm{\mu s}$ have been
found at 105~V.  This, in turn, can produce an apparent pulse height
deficit if the shaping times are not long enough.
Therefore, FAZIA extended in this experiment the memory depth for the
sampled waveforms to 
very large values (70~$\mathrm{\mu s}$) thus suppressing any ballistic
deficit. This long sampling duration was used along the measurement
for all signals and for each   voltage value.  
Some preliminary results of the data analysis (in progress) are
listed below. 
\begin{enumerate}
 \item
About energy, with very long filtering, the
non-linearities of the scale are unexpectedly small, even for
the heaviest measured ions (Kr beams were used) and even at the
lowest bias voltage (105~V). In other
words, the residual pulse height defect with respect to the full depletion
is small (within 1-2$\%$) over a wide ion and energy domain. It seems that
the supposed  dead layer (200~$\mathrm{\mu m}$ at 105V)  behaves as an active
region. This can be associated 
to the funneling phenomenon~\cite{funnel}, whose
features are rather unexplored especially for
energetic heavy ions.
 \item
According to the previous point, the capability of the telescope to
identify ions with the $\Delta E-E$ technique is essentially preserved
also in underdepleted conditions, over all the measurable spectrum
of fragments. 
\item
As expected, a huge increase of the signal timescale is observed at reduced
voltage values. In the present case, charge signal
risetimes increase  from  400~ns at 290~V to over 10000~ns at
105~V. Accordingly, the shapes of the correlations used for ion
identification (E  vs. QRT or E vs. IMAX) change a lot.

\item
About PSA we refer to the Fig.~\ref{fig-1}. 
This test confirms (Fig.~\ref{fig-1:a}) that  
good Z identification is reached at full depletion, with
threshold values  (around 50~$\mathrm{\mu m}$ for Z$\le 10$) in agreement
with previous findings~\cite{stefano,nicolas13}. Due to several not
optimized aspects (reduced preamplifier sensitivity and non excellent
doping homogeneity) only a marginal isotopic separation is observed for the
lightest fragments (up to oxygen or so) in this condition.
When decreasing the voltage, one sees an increase of the
energy limit for ion separation: at 105~V (Fig.~\ref{fig-1:b}) charges
are well separated 
above   a range roughly corresponding to the undepleted region
(200~$\mathrm{\mu m}$) of the detector. The surprising result is that
at the same threshold also a good mass separation appears for ions up
to around Neon (absent at 290~V). 
\item
Comparing the two shape variables, for underdepleted silicon, the
method E vs. IMAX offers much better separation than E 
vs. QRT  for underdepleted conditions.
\end{enumerate}

Although preliminary, these results open new possibilities. For example,
depending on the experiment, one can select the bias voltage to
tune the identification capability of the telescopes, in some cases
preferring lower thresholds with worse isotopic separation, while
privileging in other cases a good Z and A identification at the cost
of higher thresholds.  




\section{The demonstrator: construction and programs}
\begin{figure}[t]
\centering
\subfloat[]{\label{fig-2:a} \includegraphics[width=0.44\textwidth,clip]{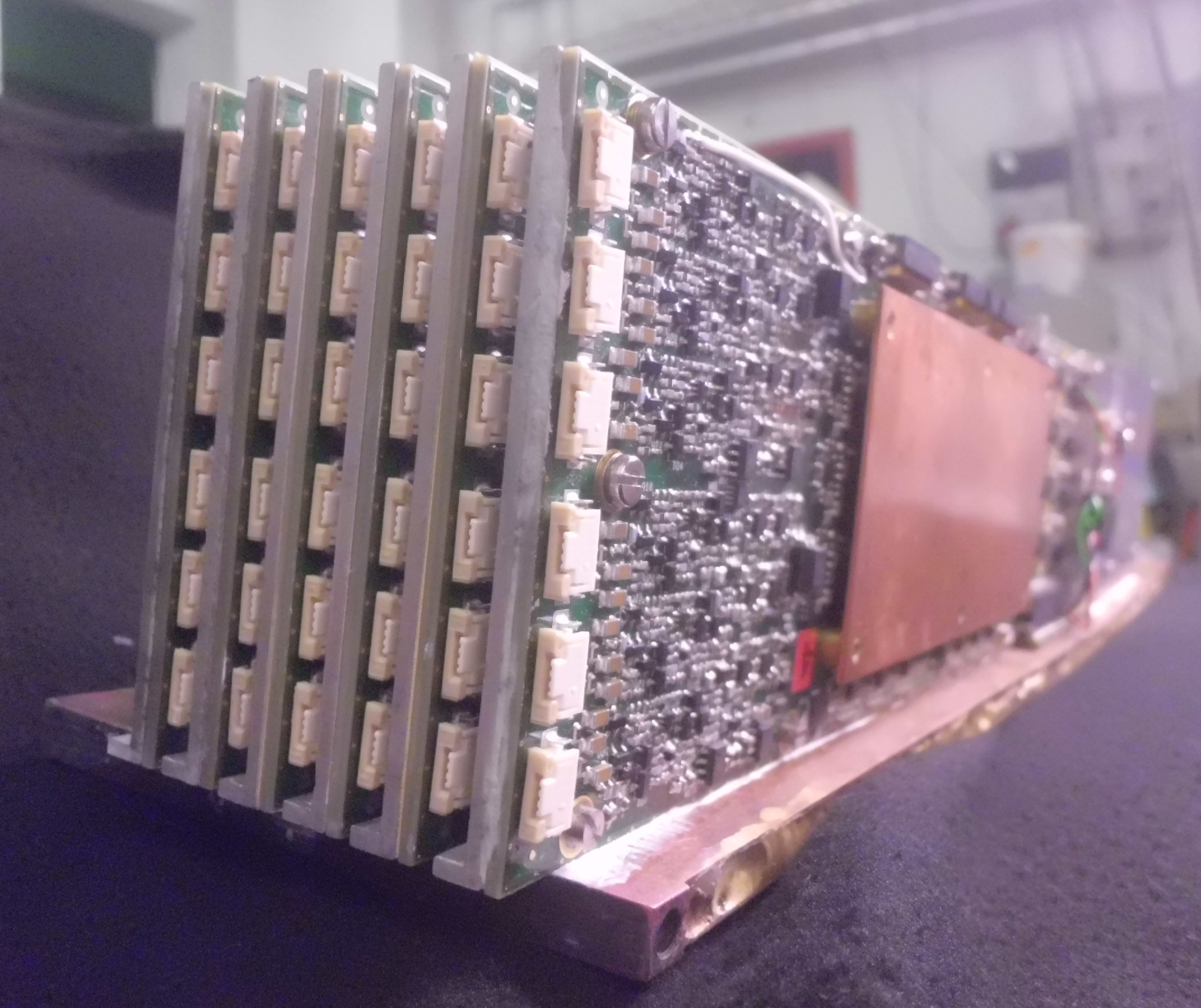}}
\subfloat[]{\label{fig-2:b} \includegraphics[width=0.5\textwidth,clip]{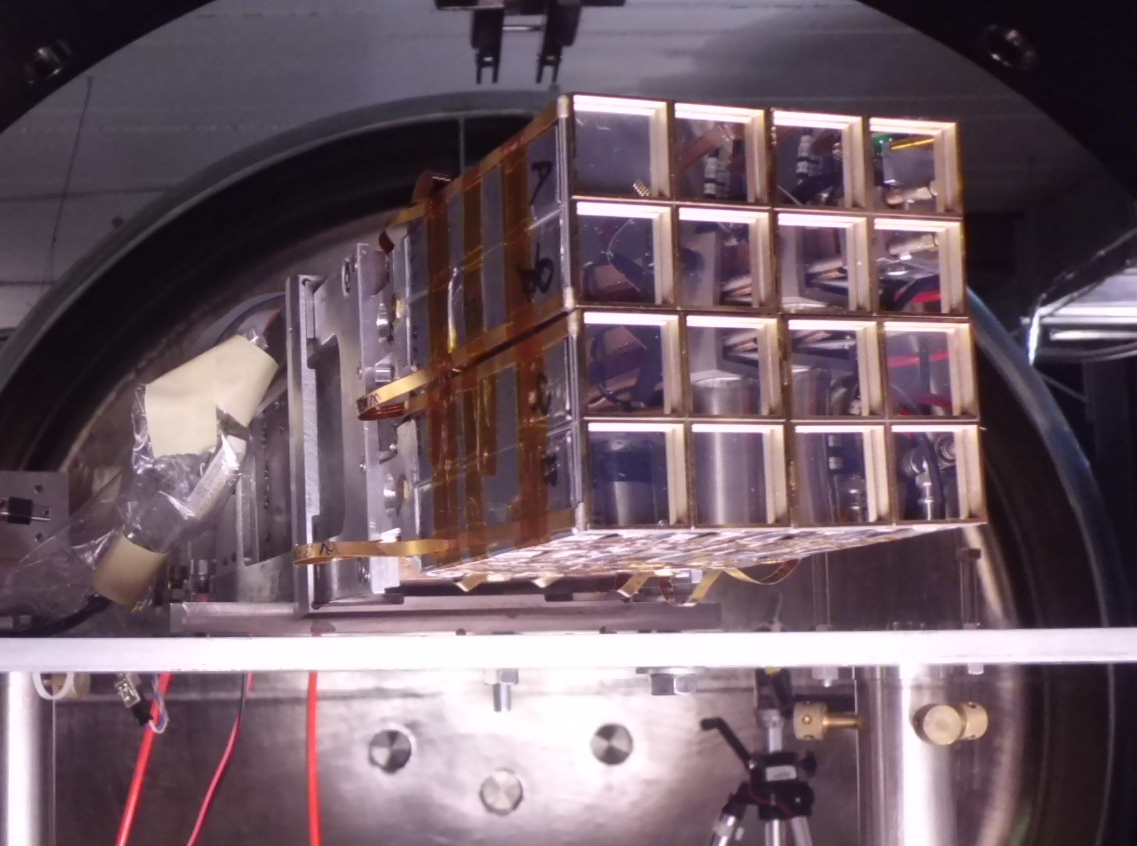}}
\caption{Left: Front view of six FEE cards mounted on the
  cooling  copper plate. The 48 female connectors will receive the
  corresponding kapton strips from the detectors. Right: Picture of
  the first FAZIA block mounted for test purposes in the Ciclope
  scattering chamber at LNS, Catania. }
\label{fig-2}       
\end{figure}

Basing on the results reported in Sect.~\ref{sec-1}), FAZIA is now
constructing an array of 192 modules organized in blocks, each
featuring 16 telescopes (see picture in Fig.~\ref{fig-2:b}). The block is a stand-alone unit and presents
original solutions to allow for vacuum operation of the FEE
electronics, located next to the detectors. In fact, 16-20~cm
long kapton strips connect the detectors (silicon and CsI) to the
preamplifiers which are hosted on specific cards together with the
ADC-FPGA stages. Each card serves
6-channels (i.e. two telescopes). Preamplifiers and digital processing
electronics are thus produced on the same card, 
reducing noise and pick-up risk and allowing for an optimum use of the FPGA
capabilities for the embedded  logics. Eight FEE cards are needed for each
block. They are 
mounted on Al shelves in a comb configuration (Fig.~\ref{fig-2:a}), tightened to a single
copper plate for efficient
cooling. Inside the plate, several channels permit the cold water flow
needed to dissipate the heat under vacuum (about 25~W per block).
On the same copper plate, other devices are mounted for many
complex functions; essentially, they transfer digitized waveforms from
electric conductors to optical fibers and perform all the slow 
operations which are remotely controlled (test pulse generation and regulation
to the preamplifiers, production, regulation of bias supply
etc. etc.).   
The blocks, from detectors to the output board,  will be supported by
special arms which allow 
several  geometries around the target, in a scattering chamber. The
foreseen mounting corresponds to a 100~cm distance of the silicon face
from the
target, but angles and block assembling can be changed at will
depending on  the experiment. The angular coverage of each block is
about $\pm 2.5 ^\circ$ and the angular resolution, defined by the
telescope dimension, is 1.2$^\circ$ at 100~cm.

FAZIA foresees the use of the demonstrator for
experiments with heavy ions at different european
laboratories~\cite{mfr12}. In the next future, isospin related 
phenomena will be studied with stable beams at intermediate
energies. Two experiments have been recently approved at the LNS; they
will be performed next year
with an array of four complete blocks (64 telescopes). The potential
of the FAZIA array in studying isospin equilibration in semiperipheral
collisions has been recently demonstrated~\cite{barli13}, although in a
simplified configuration. The reaction
products, (mainly) associated with the quasi-projectile decay, were
measured and largely identified in mass and charge for
$^{86}\mathrm{Kr}+~^{112,124}\mathrm{Sn}$ collisions at 35~MeV/u in an experiment
performed  at 
LNS. The results, reported also at this conference~\cite{piantinpc},
nicely confirm previous experimental findings~\cite{defi12} and evidence the
occurrence of both isospin drift and isospin diffusion in
semi-peripheral collisions. At the same time, our data extend
the range of ions for which isotopic variables can be studied (up to
Z=20 in this case). They  clearly show that with the FAZIA demonstrator
good quality data will be obtained to better constrain 
model parameters, especially those related to the nuclear symmetry
energy  behaviour for nuclei in  extreme conditions. 

%

\section{Conclusions}

A modular array of 192  Si-Si-CsI(Tl) telescopes is under
construction by the european FAZIA collaboration. Thanks to an
intense R$\&$D phase, the ion identification capability of such
modules has been improved both adopting specific solutions for the
detectors and using fast digital electronics channels featuring
high resolution. The silicons have been cut to present almost random
orientation and have uniform doping homogeneity. To improve PSA, they
are reverse mounted in the telescopes.  
Full charge separation has been obtained up to over Z=54 
also for ions stopping in the first silicon layer via PSA, with a
threshold of  2.5 (12) MeV/u for Carbon (Tin) ions. 
Isotopes are separated up to the iron region by the $\Delta E-E$ method for
ions stopped in second silicon or in the CsI crystal.
For slower particles, the PSA allows for isotopic
resolution up to Z$\sim$14 with 
thresholds corresponding to  $\sim $50~$ \mathrm{\mu m}$ of 
(fully depleted) silicon detector.
Preliminary results for underdepleted detectors indicate a strong improvement
of mass separation with respect to the full depletion case
at the cost of somewhat higher thresholds. \\
The expected performances of the FAZIA array will 
permit to achieve high-quality results on heavy-ion reaction dynamics,
as recently demonstrated with an experiment in which the isospin
flux has been studied at intermediate bombarding energies.

%
%

\end{document}